\DocumentMetadata{}
\documentclass[sigconf]{acmart}

\usepackage{amsmath,amsfonts}
\usepackage{algorithm}
\usepackage{adjustbox}
\usepackage{lscape}
\usepackage{siunitx}
\usepackage[noend]{algpseudocode}
\usepackage{textcomp}
\usepackage{fancyvrb}
\usepackage{graphicx}
\usepackage{soul}
\usepackage{booktabs}
\usepackage[utf8]{inputenc}
\usepackage[T1]{fontenc}
\usepackage{rotating}
\usepackage{graphicx}
\usepackage{paralist}
\usepackage{tabularx}
\usepackage{multicol}
\usepackage{multirow}
\usepackage{pbox}
\usepackage{enumitem}	
\usepackage{colortbl}
\usepackage{pifont}
\usepackage{xspace}
\usepackage{url}
\usepackage{tikz}
\usepackage{tabularx}
\usepackage{fontawesome}
\usepackage{lscape}
\usepackage{listings}
\usepackage{color}
\usepackage{showexpl}
\usepackage{anyfontsize}
\usepackage{comment}
\usepackage{soul}
\usepackage{multibib}
\usepackage{multirow}
\usepackage{xcolor}
\usepackage{xspace}
\usepackage{caption}
\usepackage{footnote}
\usepackage{tcolorbox}
\usepackage{booktabs}
\usepackage{fixltx2e}
\usepackage{balance}
\usepackage[normalem]{ulem}
\usepackage{hhline}

\newcolumntype{N}{>{\centering\arraybackslash}m{.85in}}

\AtBeginDocument{%
  }

\newboolean{showcomments}

\setboolean{showcomments}{true}

\ifthenelse{\boolean{showcomments}}
{\newcommand{\nb}[2]{
		\fbox{\bfseries\sffamily\scriptsize#1}
		{\sf\small$\blacktriangleright$\textit{#2}$\blacktriangleleft$}
	}
	
}
{\newcommand{\nb}[2]{}
	
}

\newcommand{\xmark}{\ding{55}}%
\newcommand\rev[1]{\textcolor{black}{#1}}
\newcommand\revCR[1]{\textcolor{black}{#1}}

\newcommand{\ie}{\emph{i.e.,}\xspace}
\newcommand{\eg}{\emph{e.g.,}\xspace}

\newcommand{\etal}{\emph{et~al.}\xspace}
\newcommand{\secref}[1]{Section~\ref{#1}\xspace}
\newcommand{\figref}[1]{Fig.~\ref{#1}\xspace}

\newcommand{\tabref}[1]{Table~\ref{#1}\xspace}

\newcommand{\java}{\emph{Java}\xspace}

\definecolor{lightergray}{rgb}{0.9,0.9,0.9}
\newtcolorbox{resultbox}{colback=lightergray, arc=0.5mm, top=2mm, bottom=2mm, left=2mm, right=2mm}

\copyrightyear{2024}
\acmYear{2024}
\setcopyright{rightsretained}
\acmConference[EASE 2024]{28th International Conference on Evaluation and Assessment in Software Engineering}{June 18--21, 2024}{Salerno, Italy}
\acmBooktitle{28th International Conference on Evaluation and Assessment in Software Engineering (EASE 2024), June 18--21, 2024, Salerno, Italy}\acmDOI{10.1145/3661167.3661200}
\acmISBN{979-8-4007-1701-7/24/06}

\begin{document}
\sloppy
\title{How the Training Procedure Impacts the Performance of\\Deep Learning-based Vulnerability Patching}


\author{Antonio Mastropaolo}
\affiliation{%
  \institution{Universit\`a della Svizzera italiana (USI)}
  \city{Lugano}
  \country{Switzerland}}
\email{antonio.mastropaolo@usi.ch}

\author{Vittoria Nardone}
\affiliation{%
	\institution{University of Molise}
	\city{Campobasso}
	\country{Italy}}
\email{vittoria.nardone@gmail.com}

\author{Gabriele Bavota}
\affiliation{%
	\institution{Universit\`a della Svizzera italiana (USI)}
	\city{Lugano}
	\country{Switzerland}}
\email{gabriele.bavota@usi.ch}

\author{Massimiliano Di Penta}
\affiliation{%
	\institution{University of Sannio}
	\city{Benevento}
	\country{Italy}}
\email{dipenta@unisannio.it}

\renewcommand{\shortauthors}{Mastropaolo \etal}

\begin{abstract}

Generative deep learning (DL) models have been successfully adopted for vulnerability patching. However, such models require the availability of a large dataset of patches to learn from. To overcome this issue, researchers have proposed to start from models pre-trained with general knowledge, either on the programming language or on similar tasks such as bug fixing. 
Despite the efforts in the area of automated vulnerability patching,
there is a lack of systematic studies on how these different training procedures impact the performance of DL models for such a task. This paper provides a manyfold contribution to bridge this gap, by (i) comparing existing solutions of self-supervised and supervised pre-training for vulnerability patching; and (ii) for the first time, experimenting with different kinds of prompt-tuning for this task. The study required to train/test 23 DL models.  
We found that a supervised pre-training focused on bug-fixing, while expensive in terms of data collection, substantially improves DL-based vulnerability patching. When applying prompt-tuning on top of this supervised pre-trained model, there is no significant gain in performance. Instead, prompt-tuning is an effective and cheap solution to substantially boost the performance of self-supervised pre-trained models, \ie those not relying on the bug-fixing pre-training.

\end{abstract}





\maketitle


\section{Introduction}
\label{sec:intro}
The number of reported software vulnerabilities is increasing year after year \cite{repoVul}. Such a growing trend can be explained by several factors. The most obvious is the increasing number of software projects. For example, GitHub counts at date over 370M repositories \cite{githubWiki}. \revCR{Additionally, since software projects often incorporate third-party libraries and components, they are frequently impacted by bugs. Security vulnerabilities represent a specific type of bug that can severely affect various aspects of the systems, as recently demonstrated by the Log4j vulnerability \cite{apacheLog4j}.}

To aid developers, researchers proposed solutions for the automated identification \cite{li2018vuldeepecker,chakraborty2021deep,li2016vulpecker,wu2017vulnerability,russell2018automated,hin2022linevd,zhou2021finding} and patching \cite{huang2019using,avgerinos2018mayhem,lee2018memfix,musliner2015fuzzbomb,vulre2017,seqTrans,vRepairTSE,vulRepairFSE} of vulnerabilities. For the latter task, Deep Learning (DL) models achieved state-of-the-art results thanks to their ability to infer fixing patterns by learning from concrete examples of vulnerability fixes that can be mined from software repositories. However, as highlighted by Chen \etal \cite{vRepairTSE}, a major obstacle to the adoption of DL for this task is the lack of training data. Indeed, when it comes to generative tasks (such as \emph{generating} the patched code), DL models require large supervised datasets: In this case, thousands of examples of vulnerable and fixed code. 

One way to overcome the lack of data is to rely on pre-trained models. These models are trained in two steps, namely pre-training and fine-tuning. 




In the context of vulnerability patching, Chi \etal \cite{seqTrans} and Chen \etal \cite{vRepairTSE} exploited the idea of supervised pre-training, using the task of fixing generic bugs as a pre-training objective. They showed that providing the model with such a ``bug-fixing knowledge'' before specializing it in vulnerability patching helps in boosting performance. More recently, Fu \etal \cite{vulRepairFSE} showed instead that such a supervised pre-training can be beaten by a large-scale self-supervised pre-training based on the \emph{masked language model} objective.

Besides pre-training, another strategy to overcome the lack of training data is prompt-tuning \cite{han2022ptr,lester2021power,li2021prefix,liu2021p,clark2020electra,radford2019language,raffel2020exploring,beltagy2020longformer}, namely the idea of transforming the training objective of the downstream task (\ie fine-tuning) into one that resembles the pre-training stage, and, therefore could potentially re-use information acquired during such a phase. For the vulnerability fix task, a classic fine-tuning would provide the model with a vulnerable code as input asking it to generate its patched version. A possible prompt-tuning could, for example, provide the model with the following input: \\ \textbf{Generate a patch for this vulnerable code   \{\texttt{$vulnerable_{code}$}\} as follows:  \{\texttt{$patch$}\}} where \emph{\{\texttt{$vulnerable_{code}$}\}} and \emph{\{\texttt{$patch$}\}} are placeholders to fill in each training instance. 


%


In summary, there are works suggesting the benefits of a supervised pre-training for vulnerability patching \cite{seqTrans,vRepairTSE}, others reporting the superiority of a self-supervised pre-training \cite{vulRepairFSE}, and very recent work that experimented with prompt-tuning for other software engineering tasks \cite{wang2022no}. However, there is a lack of a systematic study comparing and contrasting these training strategies (and their combinations) for the task of vulnerability patching.


This paper aims to bridge this gap, by conducting a large experimentation on the different training strategies that can be used in the context of vulnerability patch generation.
To pose the basis for the study, we first replicated existing state-of-the-art work on DL-based vulnerability patching, namely  the work by Fu \etal \cite{vulRepairFSE} which showed the superiority of self-supervised pre-training with respect to a specialized (supervised) bug-fixing pre-training \cite{vRepairTSE}. \rev{While conducting such a replication, we uncovered the existence of ``token-by-token'' duplicates within the training and the test set employed by Fu \etal \cite{vulRepairFSE}, implying a possible inflation of the reported performance. We thus cleaned the dataset, re-trained the model, and observed a substantial drop in performance as compared to what was reported by Fu \etal  \cite{vulRepairFSE}. }We then experimented on the same cleaned dataset with several different training procedures including (i) lack of pre-training, \ie the model is directly fine-tuned for the task of vulnerability patching; (ii) self-supervised pre-training and standard fine-tuning, representative of the replicate work \cite{vulRepairFSE};
(iii) supervised pre-training based on bug-fixing and standard fine-tuning, representative of previous works proposing such an idea \cite{seqTrans,vRepairTSE} and; (iv) ten different types of prompt fine-tuning were performed on top of the pre-trained models (both those with self-supervised and supervised pre-training strategies). 

Overall, our study required the training and testing of 23 models. Our main findings show that: (i) given the scarcity in fine-tuning data (\ie real-world vulnerability patches), pre-training is always beneficial for the task of vulnerability patching, independently of whether it is self-supervised or supervised; (ii) unsurprisingly, supervised pre-training (with bug fixes) is superior to the self-supervised one, contradicting previous findings \cite{vulRepairFSE}; (iii) in the context of vulnerability patching, prompt fine-tuning is a cheap mechanism to substantially boost the performance of self-supervised pre-trained models, while it does not really help models that benefited of supervised pre-training as they ``lost knowledge'' about the natural language on which they were pre-trained.

\section{Related Work}
\label{sec:related}
In this section, we discuss related work about automated vulnerability patching, as well as background notions on prompt-tuning.

\subsection{Learning-Based Vulnerability Fixing}
Researchers proposed several approaches to automatically generate fixes for software vulnerabilities. Some of these approaches are not based on machine learning, but rather leverage other techniques, such as 
property-based approaches \cite{huang2019using,lin2007autopag}, formal methods \cite{avgerinos2018mayhem,musliner2015fuzzbomb}, static analysis \cite{lee2018memfix,gao2015safe} and dynamic analysis \cite{perkins2009automatically,wang2014diagnosis}.
Given the focus of our work, we narrow down the discussion of these to ML-based approaches for vulnerability fixing.

Ma \etal \cite{vulre2017} proposed VuRLE (\ie Vulnerability Repair by Learning from Examples), an automatic tool for vulnerability detection and repair that leverages machine learning to identify and fix software vulnerabilities. VuRLE consists of a learning component and a repair component. The former collects a dataset of vulnerable and non-vulnerable code examples to train a shallow machine learning model on various features, such as code structure, control flow, and data flow. The latter chooses the most appropriate template to fix the detected vulnerability. While VuRLE can handle a wide range of vulnerabilities, it has been evaluated on a small dataset of 279 vulnerabilities.

Harer \etal \cite{Harer18} used Generative Adversarial Networks (GANs) \cite{goodfellow2020generative} to automatically repair software vulnerabilities. Both the generator and the discriminator network use a standard Neural Machine Translation (NMT) model. However, while the generator tries to generate patches, the discriminator network attempts to distinguish the crafted from the real ones. 

Chi \etal \cite{seqTrans} and Chen \etal \cite{vRepairTSE} proposed a neural Sequence-to-Sequence (Seq2Seq) approach for vulnerability patching. Both approaches are pre-trained on a bug-fix corpus and then fine-tuned on a real-world vulnerability fix dataset. The intuition behind these techniques is that the knowledge gained by the model when fixing bugs (supervised pre-training) can be transferred to the generation of patches for vulnerable code. 
Although both approaches use a neural Sequence-to-Sequence (Seq2Seq) learning model to generate fixed code sequences, they differ in the programming language they handle (\java \emph{vs} C) and the type of vulnerabilities the model can patch (inline \emph{vs} multi-line).


To the best of our knowledge, the latest technique proposed for vulnerability patching is VulRepair, introduced by Fu \etal \cite{vulRepairFSE}. This approach leverages the T5 model introduced by Raffel \etal~\cite{raffel2020exploring}.
Rather than starting from randomly initialized weights, Fu \etal decided to build VulRepair employing CodeT5 as a pre-trained model. This design choice reduces the burden of pre-training a new T5 model from scratch and leverages the condensed knowledge of the model. VulRepair has been evaluated on two different datasets (CVEFixes \cite{bhandari2021cvefixes} and Big-Vul \cite{fan2020ac}) which consist of 8,482 pairs of vulnerability fixes in total,
and for which Fu \etal reported an ability of the technique to generate patches as a real human would do in up to 44\% of cases.



\subsection{Prompt-tuning, and its applications to software engineering tasks}

\label{sub:prompt}


Prompt-tuning is a technique aimed at transforming the training objective of downstream tasks to resemble the task performed during the pre-training stage, which utilizes the MLM (Masking Language Model) objective \cite{devlin2018bert,liu2019roberta}. This is done by modifying the model input. That is, if the pre-training is carried out on natural language,  the idea is to incorporate a natural language prompt (\eg an English-written sentence), resulting in an input format consistent with the pre-training stage. We can distinguish between two types of prompt-tuning, depending on how the prompt is injected.

\subsubsection{Hard Prompt:} The hard prompt \cite{han2022ptr,gu2021ppt,schick2020exploiting} instruments the model input through the addition of fixed natural language elements, known as prompts. The goal is to leverage task-specific knowledge acquired during the pre-training. This type of prompt is considered ``hard'' because it consists of discrete tokens that have a clear meaning and can be understood by a human.
To generate hard templates, we re-arranged the model's input tokens according to the designed prompt (\secref{sec:techniques})

\subsubsection{Soft Prompt:}  soft prompt techniques \cite{han2022ptr,li2021prefix,tsimpoukelli2021multimodal} do not use natural language words as a template but, rather, continuous vectors learned when specializing the model (\ie fine-tuning) on the downstream task(s).
Soft prompt techniques can be distinguished into \emph{vanilla soft prompt} and \emph{prefix soft prompt}.
The former replaces the hard prompt tokens with virtual tokens. The latter arranges several virtual tokens at the beginning of the input. For both prompting techniques, the embeddings of these virtual tokens are learned during the fine-tuning phase of the model.  
For generating vanilla soft prompts, we replace the natural language tokens in the hard prompt templates with virtual tokens. 


\subsubsection{Applications of prompting to software engineering tasks:}
Wang \etal \cite{wang2022no} postulated the possibility of leveraging prompt- tuning for software engineering tasks. Specifically, they start from pre-trained CodeBERT \cite{feng2020codebert} and CodeT5 \cite{wang2021codet5} models, and experiment with three software engineering tasks, namely defect prediction, code summarization, and code translation. Their results indicate that prompt-tuning: (i) always outperforms fine-tuning, (ii) it is particularly useful when fine-tuning material is limited, and (iii) its performances depend on the prompt length.

Nashid \etal \cite{nashidretrieval} proposed CEDAR, a framework to automatically build effective prompts for two code-related tasks, \ie test assertion generation and program repair. CEDAR selects prompts by applying an automated retrieval-based demonstration selection strategy. CEDAR takes as input a set of code demonstrations and produces as output a text-based prompt. Nashid \etal \cite{nashidretrieval} showed that CEDAR outperforms the best-performing state-of-the-art models for code-related tasks, like assertion generation and program repair. Furthermore, Nashid \etal \cite{nashidretrieval} showed that prompt-tuning can achieve high accuracy in specific downstream tasks. Leveraging the experience of Wang \etal \cite{wang2022no} and Nashid \etal \cite{nashidretrieval}, we empirically evaluate the impact of different training strategies aiming at generating vulnerability patches, by also experimenting with ten different types of prompt fine-tuning.


\section{Study Definition, Design, and Planning}
\label{sec:context}

The \emph{goal} of this study is to evaluate how the use of various training procedures affects the effectiveness of DL-based models in patching code vulnerabilities. We investigate combinations of different pre-trainings and fine-tunings, as well as the use of different types of hard and soft prompt-tuning. The \emph{context} consists of a Code-pretrained T5 model \cite{wang2021codet5}, a bug-fixing dataset from Chen \etal~\cite{vRepairTSE}, and a vulnerability patches dataset made available with the VulRepair \cite{vulRepairFSE} paper, after the removal of duplicates, as explained in \secref{sec:ft-dataset}.

\subsection{Research Questions}
\label{sec:rq}
We address the following research question (RQ):
\begin{quote}
\textbf{RQ$_1$:} \textit{How do different training procedures impact the performance of DL-based vulnerability patching systems?}
\end{quote}

To address this general research question, we formulate three sub-questions, each one aimed at studying in isolation the impact on performance of a specific training technique:

\begin{compactitem}
	
	\item[\textbf{RQ$_{1.1}$:}]\textit{To what extent does the usage of a self-supervised pre-training objective benefit the generation of vulnerability patches?} 
	In RQ${1.1}$ we assess the impact of the classic \emph{masked language model} self-supervised objective on the performance of a DL-based technique fine-tuned for vulnerability patching. 
	
	\item[\textbf{RQ$_{1.2}$:}]\textit{To what extent does the usage of a supervised pre-training objective benefit the generation of vulnerability patches?} RQ$_{1.2}$ evaluates the effectiveness of further pre-training an already (self-supervised) pre-trained model using a supervised objective resembling the downstream task. 
	
	In our case, we use as a supervised pre-training objective the task of fixing bugs as a more general representation of vulnerability patching.
	
	\item[\textbf{RQ$_{1.3}$:}]\textit{What is the role played by prompt-tuning when patching vulnerable code?} RQ$_{1.3}$ investigates the impact on the performance of a prompt fine-tuning (as compared to a standard fine-tuning) by comparing the use of the hard prompt (for which we experiment with different natural language templates) and soft prompt.

\end{compactitem}

%

\subsection{Context: Datasets} \label{sec:datasets}
We describe the pre-training and fine-tuning datasets we leverage in our study. \secref{sec:techniques} will detail how they have been used to train different models aimed at assessing the impact of the different training strategies. We anticipate that the fine-tuning dataset used to train the models for the task of vulnerability patching includes C/C++ functions, thus explaining the focus on these languages for the other datasets as well.

\subsubsection{Self-supervised pre-training}
When experimenting with self-supervised pre-training using the \emph{masked language model} objective (\ie masking 15\% of tokens in a string and asking the model to predict them), we exploit CodeT5 \cite{wang2021codet5}, a Text-To-Text Transfer Transformer (T5) model \cite{raffel2020exploring} already pre-trained on code and widely used in the software engineering literature. Thus, we describe in the following the dataset that has been used for its pre-training. 

Wang \etal \cite{wang2021codet5} leveraged the CodeSearchNet dataset \cite{husain2019codesearchnet} to pre-train CodeT5. This dataset comprises both technical natural language (\ie code comments) and code. Additionally, Wang \etal collected supplementary data from C/C\# repositories on GitHub. This resulted in a total of 8,347,634 pre-training instances, where an instance is a code function: 3,158,313 of these functions are coupled with their documentation, while 5,189,321 feature only code.

\subsubsection{Supervised pre-training on bug-fixing}
\label{sec:ft-dataset-bugs}
Chen \etal \cite{vRepairTSE} proposed the use of a supervised pre-training based on bug-fixing before fine-tuning the model for the task of vulnerability patching. We thus use the bug-fixing corpora they provide as a supervised pre-training dataset. To build this dataset, the authors mined 729M GitHub commits identifying 21M of them as bug fixes. Out of these, 910k C/C++ bug-fixing commits were selected. Then, function-level changes performed to fix the bug were collected as $\langle F_b, F_f \rangle$ pairs, where $F_b$ and $F_f$ represent the buggy and the fixed version, respectively, of a function $F$. After duplicate removal, the authors were left with 544,858 valid instances which can be used to train a model for the bug-fixing task (\ie input $F_b$, expected output $F_f$). Out of these pairs, 534,858 are used for training, and the remaining 10,000 pairs for validation (\ie to identify the best-performing pre-trained model for the task of bug-fixing).

\subsubsection{Fine-tuning for vulnerability patching}
\label{sec:ft-dataset}

We reuse the dataset provided in the replication package of VulRepair (Fu \etal \cite{replication-vulrepair}) under the path \texttt{data/fine\_tune\_data}. This dataset contains 8,482 vulnerability fixes (a pair of vulnerable C/C++ functions and their patched version) obtained by merging two datasets: CVE-Fixes \cite{bhandari2021cvefixes} and Big-Vul \cite{fan2020ac} and then split into 70\% for training (5,937), 20\% (1,706) for testing and 10\% (839) for validation.

Fu \etal \cite{vulRepairFSE} processed these instances to help the model learning. Such a procedure starts with marking the vulnerable code snippet (\ie the input sequence) using two special tags: $\langle StartLoc \rangle$ and $\langle EndLoc \rangle$, where $\langle StartLoc \rangle$ tags the beginning of the vulnerable code lines, while $\langle EndLoc \rangle$ identifies their end.
Two additional special tokens $\langle ModStart \rangle$ and $\langle ModEnd \rangle$ indicate the beginning and end, respectively, of the patching code (\ie the output sequence). Using these tags the model is trained to pay more attention to the code elements involved in the vulnerability patching. 

\rev{Upon examining the dataset used by Fu \etal \cite{vulRepairFSE}, we found duplicated instances among the training, validation, and test sets. The presence of such overlapping instances would pose questions about the real ability of the model to patch vulnerabilities. This, in turn, could lead to artificially boosted performance metrics.} Thus, to address this issue we conducted a systematic analysis to identify problematic instances.

We found and discarded 38 instances (\ie pairs of $\langle vulnerable_{code}$, $patch \rangle$) having an empty string as $patch$ (24 in the training set, 3 in the validation, and 11 in the test). After this preliminary cleaning, we addressed the issue of overlapping instances across the three sets. We removed from the training set all samples having an identical counterpart in the other (validation or test) sets. In particular, we removed from the training set 674 instances shared with the test set, and 322 shared with the validation set. We decided to remove such instances from the training rather than from the test/validation set because we wanted to evaluate the impact on VulRepair's performance when being fine-tuned on the cleaned dataset by comparing our findings with those reported in the original paper \cite{vulRepairFSE}. To do this, it was important to exploit exactly the same test set which, thus, should stay unchanged.

The number of instances we removed during this deduplication step is higher (1,008) than the sum of overlapping instances between the training and  both the test and validation sets (674 + 322 = 996). This is because the training set itself contained some duplicates and, thus, some instances removed as shared with the test/validation set have been removed more than once since appearing multiple times in the training set.

\begin{table}[h!]
	\centering
	\caption{Number of instances included in the dataset used by Fu \etal \cite{vulRepairFSE}, before and after performing the cleaning}
		\label{tab:dataset}
			\resizebox{0.7\columnwidth}{!}{%
		\begin{tabular}{lrrrr}
			\hline
			\vspace{0.03cm}
			\textbf{Dataset}  & \textbf{Train} & \textbf{Test} & \textbf{Eval}  & \textbf{Overall}\\
			\hline
			
			VulRepair \cite{vulRepairFSE} & 5,937 & 1,706 & 839 & 8,482\\
			\midrule
			$VulRepair_{WET}$  & 5,913 & 1,695 & 836 & 8,444\\
				\midrule
			$VulRepair_{FC}$  & 4,905 & 1,695 & 836 & 7,436\\
			\bottomrule
		\end{tabular}
	}
\scriptsize
\vspace{-3mm}
\end{table}

\tabref{tab:dataset} reports the number of instances included in the fine-tuning dataset for each split (\ie train, validation, and test) before and after performing the above-described cleaning. 

The first row reports the number of instances included in the original dataset (\ie before cleaning), while the second and third rows show $VulRepair_{WET}$ (Without Empty Targets) and $VulRepair_{FC}$ (Fully-Cleaned), respectively, with $VulRepair_{FC}$ being the one used in our study. 

\subsection{Context: Experimented Training Strategies}
\label{sec:techniques}


We detail the training strategies we experiment with as follows:

\subsubsection{No Pre-training + Fine-tuning}
We fine-tune on the $VulRepair_{FC}$ dataset a T5$_{base}$ model \cite{raffel2020exploring} (\ie the same used for CodeT5 \cite{wang2021codet5}) without any pre-training. Such a model will serve as a baseline to assess the impact of pre-training on the model's performance. 

\subsubsection{Self-supervised Pre-training + Fine-tuning}
\label{sub:vulrepair}
We replicate the VulRepair approach by Fu \etal \cite{vulRepairFSE} by, however, fine-tuning the model on $VulRepair_{FC}$ (\ie the cleaned version of their dataset). VulRepair exploits the pre-trained CodeT5 which, as previously said, used a self-supervised \emph{masked language model} pre-training objective. Such a model will allow us to answer RQ$_{1.1}$.

\subsubsection{Self-supervised \& Supervised Pre-training + Fine-tuning}
\label{sub:bugfixmodel}
Our goal here is to train a model taking advantage of both self-supervised (\emph{masked language model}) and supervised (\emph{bug-fixing}) pre-training before being fine-tuned for vulnerability patching. Such an approach is inspired by the works of Chi \etal \cite{seqTrans} and Chen \etal \cite{vRepairTSE}. We start again from the (self-supervised) pre-trained CodeT5. Then, we further train it for 5 epochs using the bug-fixing dataset described in \secref{sec:ft-dataset-bugs}. Finally, we specialize the model to the downstream task using the $VulRepair_{FC}$ dataset.

\subsubsection{Prompt Fine-tuning}
\label{sub:prompt-design}
As previously mentioned with prompt fine-tuning each fine-tuning instance undergoes a transformation  shaping it so that it resembles the data encountered during the pre-training procedure. We experiment prompt fine-tuning on top of the pre-trained models described in Sections \ref{sub:vulrepair} and \ref{sub:bugfixmodel}. This means that, instead of performing a conventional fine-tuning procedure, we replaced it with hard or soft prompt fine-tuning. 

For both soft and hard prompting, we experimented with five different templates that can be found in our online appendix~\cite{replication}.

The prompt templates embed the $vulnerable_{code}$ and the $patch$ in a natural language sentence which could either be a very simple sentence describing the need to patch the code, or a more verbose one, also specifying information extracted from CWE \cite{martin2008common} about the vulnerability affecting $vulnerable_{code}$. An example is ``\emph{Exposure of Sensitive Information to an Unauthorized Actor}'' related to CWE id number $200$, which draws out a weakness of code to expose sensitive information to a not authorized actor gaining access to that. 



To collect textual information of CWE, \ie name and description, we downloaded from the Mitre website \cite{cweSite} the CSV file containing the CWE List with all related information. Then, given the CWE id contained in the dataset of Fu \etal~\cite{vulRepairFSE} we used it to search into the CSV file for the corresponding CWE name and description. Note that 12 CWE ids were not contained in the CSV file since these ids referred to CWE categories and not to a specific weakness, \eg CWE-399. Thus, for these 12 CWE ids, we searched directly on the Mitre website \cite{cweSite} to retrieve their name and description.

We use the OpenPrompt library \cite{ding2021openprompt} to carry out soft-prompting (\ie to learn the needed continuous vectors --- see \secref{sub:prompt}) as done by Wang \etal \cite{wang2022no}.

Overall, we prompt fine-tuned ten versions (5 with hard and 5 with soft prompting) of the self-supervised pre-trained model described in \secref{sub:vulrepair} and ten of the self-supervised \& supervised pre-trained model from \secref{sub:bugfixmodel}.

\begin{table*}[ht!]
	\caption{Exact Match (\ie the recommended code is equal to the oracle) and CrystalBleu scores achieved by the different techniques when patching vulnerable C functions. In dark-grey boxes we report the highest value for both metrics when producing $K$=1, $K$=3, $K$=5, and $K$=10 candidate patches. In \texttt{boldface} the best prompt fine-tuning template within each pre-training strategy.}
	\label{tab:perfect}
	\centering
	\resizebox{0.85\textwidth}{!}{%
\begin{tabular}{lll>{\columncolor[gray]{0.8}}rr>{\columncolor[gray]{0.8}}rr>{\columncolor[gray]{0.8}}rr>{\columncolor[gray]{0.8}}rr}
	\hline
		\multirow{2}{*}{{\bf Model-ID}} &  \multicolumn{2}{c}{{\bf Training Procedure}} 
		&   \multicolumn{2}{N}{\bf Top-1} 
		&  \multicolumn{2}{N}{\bf Top-3}  
		&  \multicolumn{2}{N}{\bf Top-5} 
		&   \multicolumn{2}{N}{\bf Top-10} \\    \hhline{~----------}
	   & {\bf Pre-training} & {\bf Fine-tuning} 
	   &  {\bf EM} & {\bf CB}  
	   &  {\bf EM} &  {\bf CB}
	   & {\bf EM} &  {\bf CB}
	   & {\bf EM} & {\bf CB} \\
	   \hline
		
	M0 & \hspace{0.75cm} \xmark	& Supervised & 2.35\%   & 31.62\%               & 3.54\%   & 32.66\%            &  4.18\%   & 32.34\%       &   4.79\% &  31.70\%   \\

	\hline
	M1 \cite{vulRepairFSE} & Self-supervised & Supervised &   3.34\%   & 40.18\%             & 5.72\%     & 42.10\%         & 6.54\%   & 41.14\%        & 6.90\%  & 38.58\%         \\
		
		\hline
	M2 & Self-supervised + Supervised & Supervised & 12.28\%   & 47.98\%              &  17.58\%   & 50.21\%          & 18.64\%  & 49.61\%         & 18.82\%     & 47.26\%       \\
	
	\hline
	$M3_{S1}$ & Self-supervised & Soft prompt-tuning &  5.48\%        & 50.52\%        		& 7.02\%        	 		& 51.53\%        		 			&  7.02\%       		  &  51.22\%      &   7.13\%        		& 50.80\%       \\
	$M3_{S2}$ & Self-supervised & Soft prompt-tuning &  6.43\%        & 50.96\%       	    				  &  \textbf{7.43\%}    & 52.53\%      					 &  \textbf{7.55\%}   & 51.92\%       &  \textbf{7.70\%}   		& 51.37\%      \\
	$M3_{S3}$ & Self-supervised & Soft prompt-tuning & 6.01\%         & 50.00\%        		 			  &  7.13\%        	 		 & 51.32\%         				 &  7.31\%      			 & 50.71\%        &  7.37\%       		& 49.64\%      \\
	$M3_{S4}$ & Self-supervised & Soft prompt-tuning &  6.66\%       & \textbf{52.51\%}       & 7.31\%     			   &  \textbf{53.35\%}     	 &  7.37\%        		   & \textbf{52.48\%}       &  7.55\%       &  \textbf{51.95\%}      \\
	$M3_{S5}$ & Self-supervised & Soft prompt-tuning &  \textbf{6.78\%}  & 51.35\%   	   			   &  7.37\%        		  &   52.11\%        	    	& 7.43\%         & 52.04\%       & 7.37\%       		& 51.50\%      \\
	
	\hline
		$M3_{H1}$ & Self-supervised & Hard prompt-tuning  & 3.83\%          	   & 40.24\%        	    		& 5.60\%      		& 41.59\%        				& 6.37\%         & 41.13\%         & 6.78\%     	& 38.36\%      \\
    $M3_{H2}$ & Self-supervised &  Hard prompt-tuning & 3.60\%         		  & \textbf{40.96\%}         		  &  5.48\%     	  & 41.78\%           			     &  6.49\%        & 41.20\%         &  6.72\%   	 & 38.54\%      \\
	$M3_{H3}$ & Self-supervised & Hard prompt-tuning  & 3.71\%           	   & 40.75\%           &    5.36\%   	   & 41.71\%     				   &   6.25\%       & 40.81\%         &  6.54\%   	   & 38.54\%     \\
	$M3_{H4}$ & Self-supervised &  Hard prompt-tuning &  \textbf{4.01\%}   & 40.69\%            		 &    \textbf{6.13\%}    & 41.95\%    		  &   \textbf{7.02\%}  & 40.92\% 		& \textbf{7.55\%}   & 39.04\%       \\
    $M3_{H5}$ & Self-supervised &  Hard prompt-tuning &  3.60\%         	 & 39.77\%        		&  5.78\%            & \textbf{42.00\%}      	     &  6.32\%        & \textbf{41.30\%}    &  \textbf{7.55\%}  & \textbf{39.80\%}       \\

   \hline
  $M4_{S1}$ & Self-supervised + Supervised & Soft prompt-tuning &   8.96\%       		  & \cellcolor[HTML]{656565}\color[HTML]{FFFFFF} \bf 55.28\%         &  10.67\%    	&  \cellcolor[HTML]{656565}\color[HTML]{FFFFFF} \bf 56.13\%     		 &   10.97\%       & 55.87\%          & 10.91\%   		&  55.12\%       \\
  $M4_{S2}$ & Self-supervised + Supervised & Soft prompt-tuning &  \textbf{9.43\%}    & 54.45\%         &  10.85\%       & 56.11\%        &   11.15\%        & 55.83\%          &   11.26\%  	 & \cellcolor[HTML]{656565}\color[HTML]{FFFFFF} \bf 55.15\%      \\
  $M4_{S3}$ & Self-supervised + Supervised & Soft prompt-tuning &  8.67\%       	 	 & 53.86\%           &   \textbf{11.10\%}  	   & 55.27\%      	    &    11.15\%        & 54.77\%      	&   11.26\%  	 &  53.55\%     \\
  $M4_{S4}$ & Self-supervised + Supervised & Soft prompt-tuning &  8.90\%       		 & 54.13\%          &   11.03\% 			& 55.56\%       		&    \textbf{11.56\%}  & \cellcolor[HTML]{656565}\color[HTML]{FFFFFF} \bf 55.89\%    &  \textbf{11.62\%}  &  54.34\%      \\
  $M4_{S5}$ & Self-supervised + Supervised & Soft prompt-tuning &  8.84\%       		 & 53.77\%          &     10.73\% 		  & 55.32\%  		&      11.10\%     & 55.04\%    		&     11.26\%  	 &  53.75\%    \\
   
   \hline
  $M4_{H1}$ & Self-supervised + Supervised & Hard prompt-tuning & \cellcolor[HTML]{656565}\color[HTML]{FFFFFF} \bf 13.21\%   & \textbf{49.38\%}        &  \cellcolor[HTML]{656565}\color[HTML]{FFFFFF} \bf 18.29\%              & \textbf{51.82\%}       & \cellcolor[HTML]{656565}\color[HTML]{FFFFFF} \bf 19.46\%   & \textbf{50.86\%}        &  \cellcolor[HTML]{656565}\color[HTML]{FFFFFF} \bf 20.11\%  & \textbf{48.67\%}       \\
  $M4_{H2}$ & Self-supervised + Supervised & Hard prompt-tuning & 12.33\%          & 48.26\%      &  17.34\%        			& 50.31\%          &  18.70\%       & 50.23\%        &  19.29\%  & 47.77\%       \\
  $M4_{H3}$ & Self-supervised + Supervised & Hard prompt-tuning &   12.33\%           	& 50.61\%          & 17.70\%            & 50.65\%       & 18.93\%   		& 50.40\%        & 19.58\%   & 47.64\%       \\
  $M4_{H4}$ & Self-supervised + Supervised & Hard prompt-tuning & 11.97\%           & 48.04\%        & 17.52\%         			& 51.13\%        &  18.87\%   		& 50.23\%       & 19.58\%  & 48.16\%         \\
  $M4_{H5}$ & Self-supervised + Supervised & Hard prompt-tuning & 12.03\%          & 48.15\%        &  17.46\%       				 & 50.81\%         &   18.05\%  			& 49.48\%       & 18.76\%    & 47.17\%      \\
   \hline
\end{tabular}
}
\end{table*}

\subsection{Data Collection and Analysis}
\label{sec: data-collection}
For all trained models, the fine-tuning (or prompt fine-tuning) performed on the $VulRepair_{FC}$ dataset has been run for 75 epochs, in line with what was done in previous work \cite{vulRepairFSE}. After each epoch, we assess the loss of the model on the evaluation set, selecting as the best checkpoint the one having the lowest loss. This is the checkpoint that has then been evaluated by running it on the test set. In total, we trained and tested 23 models: 

\begin{itemize}
\item 1: \emph{No pre-training + fine-tuning}, in the results referred as M0;

\item 1: \emph{Self-supervised pre-training + fine-tuning}, referred as M1;

\item 1: \emph{Self-supervised \& supervised pre-training + fine-tuning}, referred as M2;

\item 10: \emph{Self-supervised pre-training + prompt fine-tuning} $\times$ 10 prompt templates (5 for soft, referred as $M3_{S1-5}$, and 5 for hard prompting, referred as $M3_{H1-5}$, where the last digit 1-5 indicates the used prompt among those described in \secref{sub:prompt-design}). For example, $M3_{S1}$ uses soft prompting and the first template;

\item 10: \emph{Self-supervised \& supervised pre-training + prompt fine-tuning} $\times$ 10 prompt templates (5 for soft, referred as $M4_{S1-5}$, and 5 for hard prompting, referred as $M4_{H1-5}$).

\end{itemize}

We run each trained model on the 1,695 functions in the test set, asking it to generate patches. We use the beam search decoding schema \cite{freitag2017beam} that allows to produce multiple vulnerability repair candidates for an input sequence. We assess the performance of each model using two metrics: (i) the percentage of Exact Match (EM) predictions for different beam sizes $K$ (EM@K), and (ii) the CrystalBLEU score \cite{eghbali2022crystalbleu}.

\textbf{EM@K} measures the percentage of instances in the test set for which the  sequence predicted by the model matches the expected oracle sequence. 

Since we use beam-search, we report the results for different values of $K$ (\ie 1, 2, 3, 4, 5, 10), as done in \cite{vulRepairFSE}. We avoid reporting results with $K$=50, because, as pointed out by Fu \etal \cite{vulRepairFSE}, the effort security analysts have to put into manually inspecting such a large number of patches may hinder the adoption of the approach in practice. 

\textbf{CrystalBLEU score} \cite{eghbali2022crystalbleu} measures how similar the candidate (predicted code) and reference code (oracle) are, similar to how the BLEU score \cite{papineni2002bleu} measures similarity between texts. However, CrystalBLEU is specifically designed for code evaluation, while retaining desirable properties of BLEU, specifically being language-agnostic and minimizing the effect of trivially shared $n$-grams, which would produce inflated results.\smallskip

Also, we perform statistical tests to determine whether one of the experimented techniques is more effective in patching the vulnerable code. To this end, we use McNemar's test \cite{mcnemar} (which is a proportion test for dependent samples) and Odds Ratios (ORs) on the EMs that the techniques can generate. We also statistically compare the distribution of the CrystalBLEU scores (computed at the sentence level) for the predictions generated by each technique by using the Wilcoxon signed-rank test \cite{wilcoxon}. The Cliff’s Delta (d) is used as effect size \cite{Cliff:2005} and it is considered: negligible for $|d|$  0.10, small for 0.10 $\le$ $|d|$ < 0.33, medium for 0.33 $\le$ $|d|$ < 0.474, and large for $|d|$ $\ge$ 0.474. For all tests, we assume a significance level of 95\% and we account for multiple tests by adjusting $p$-values using Holm’s correction procedure \cite{Holm1979a}.

To make it easier for the reader to follow the manuscript, we have assigned a unique ID to each model/configuration we tested, which can be found in \tabref{tab:perfect}.

\section{Results}
\label{sec:result}

\begin{table}[t]
\centering
\caption{Comparison among different training strategies for top-1 predictions: McNemar's and Wilcoxon's test results.}
\label{tab:stats}
\begin{adjustbox}{max width=0.8\columnwidth}
\begin{tabular}{lrr|rr}
  \hline
  \multirow{2}{*}{\textbf{Comparison}} &  \multicolumn{2}{c}{\textbf{McNemar's Test}} & \multicolumn{2}{c}{\textbf{Wilcoxon's Test}} \\ \cline{2-5}
 & \textbf{\emph{p}-value} & \textbf{OR}  & \textbf{\emph{p}-value} & \textbf{d} \\ 

\hline

M0 vs. M1 \cite{vulRepairFSE} & 0.03 & 1.81 & $<$0.05 & 0.235 (S) \\ 

 \hline 
  
  \rowcolor[gray]{.8}
 M1 \cite{vulRepairFSE} vs. M2 & $<$0.05 & \cellcolor[HTML]{656565}\color[HTML]{FFFFFF} \bf 26.33 & $<$0.05 & 0.140 (N) \\ 
   
 \hline
      \rowcolor[gray]{.9}
 M1 \cite{vulRepairFSE} vs. $M3_{S1}$& $<$0.05   & 2.51 & $<$0.05 & 0.199 (S) \\ 
    \rowcolor[gray]{.9}
 M1 \cite{vulRepairFSE} vs. $M3_{S2}$& $<$0.05  & 2.82 & $<$0.05 & 0.198 (S) \\ 
    \rowcolor[gray]{.9}
 M1 \cite{vulRepairFSE} vs. $M3_{S3}$ & $<$0.05 & 2.57 & $<$0.05 & 0.186 (S) \\ 
    \rowcolor[gray]{.9}
 M1 \cite{vulRepairFSE} vs. $M3_{S4}$ & $<$0.05 & 2.96 & $<$0.05 & 0.232 (S) \\ 
    \rowcolor[gray]{.9}
 M1 \cite{vulRepairFSE} vs. $M3_{S5}$ & $<$0.05 & \cellcolor[HTML]{656565}\color[HTML]{FFFFFF} \bf 3.03 & $<$0.05 & 0.203 (S) \\ 

\hline
  
  M1 \cite{vulRepairFSE} vs. $M3_{H1}$   &  1.0 & 1.63 & 1.0 & -0.002 (N) \\ 
  M1 \cite{vulRepairFSE} vs. $M3_{H2}$  &  1.0 & 1.42 & 0.3 & 0.018 (N) \\ 
  M1 \cite{vulRepairFSE} vs. $M3_{H3}$  &  1.0 & 1.55 & 1.0 & 0.012 (N) \\ 
  M1 \cite{vulRepairFSE} vs. $M3_{H4}$  & 0.6 & 1.83 &1.0 & 0.009 (N) \\ 
  M1 \cite{vulRepairFSE} vs. $M3_{H5}$  & 1.0 & 1.18 & 0.5 & -0.013 (N) \\

   \hline
     \rowcolor[gray]{.9}
  $M4_{S1}$ vs. M2 &  $<$0.05  & 1.98 & $<$0.05 &  	  -0.129 (N)  \\ 
    \rowcolor[gray]{.9}
  $M4_{S2}$ vs. M2 & $<$0.05  &  1.81 & $<$0.05 &     -0.114 (N) \\ 
    \rowcolor[gray]{.9}
  $M4_{S3}$ vs. M2 & $<$0.05  & \cellcolor[HTML]{656565}\color[HTML]{FFFFFF} \bf 2.36 & $<$0.05 &     -0.105 (N) \\ 
    \rowcolor[gray]{.9}
  $M4_{S4}$  vs. M2 & $<$0.05 & 2.25 & $<$0.05 &     -0.109 (N) \\ 
    \rowcolor[gray]{.9}
  $M4_{S5}$ vs. M2 & $<$0.05  & 2.09 & $<$0.05 &     -0.101 (N) \\

  \hline
  M2 vs. $M4_{H1}$ & 0.32 & 1.77 & $<$0.05 & 0.024 (N) \\ 
 $M4_{H2}$ vs. M2 & 1.0 & 1.05 & 0.60  & -0.005 (N) \\  
 $M4_{H3}$ vs. M2 & 1.0 & 1.05 & 0.60 & -0.005 (N) \\ 
 $M4_{H4}$  vs. M2 &1.0 & 1.46 & 0.60 & -0.0001 (N) \\ 
 $M4_{H5}$ vs. M2 & 1.0 & 1.37 & 0.60 & -0.003 (N) \\

  \hline
\end{tabular}
\end{adjustbox}
\vspace{-0.6cm}
\end{table}

\tabref{tab:perfect} reports the results achieved using the different training strategies subject of our study. The first column, ``Model-ID'' reports a unique identifier we assigned to each of the 23 trained models, using the notation $Mx_{(S|H)1-5}$ described in \secref{sec: data-collection}.
The ``Pre-training'' and ``Fine-tuning'' columns indicate the combination of training procedures adopted for each model. For example, the first line (\ie M0), represents the non-pre-trained model which has been directly fine-tuned for the task of vulnerability patching using a standard fine-tuning procedure. 
Finally, EM and CB report the performance achieved by a specific configuration in terms of (i) the percentage of predictions being Exact Matches (EM), and (ii) the average CrystalBLEU score \cite{eghbali2022crystalbleu} across all predictions in the test set (CB). To enhance the readability of such a table and to ease the results' discussion, we only include in the manuscripts the results achieved with $K$ (\ie the beam size) of 1, 3, 5, and 10. The comprehensive set of results with all beam sizes is available in our replication package \cite{replication}.

\tabref{tab:stats} reports the results of the statistical tests (Fisher's exact test and Wilcoxon signed-rank test), with adjusted $p$-values, OR, and Cliff's $d$ effect size. An $OR>1$, or a positive Cliff's $d$ indicates that the right-side treatment outperforms the left-side one.  For the sake of readability, we ordered the treatments to show ORs that are generally $\geq 1$.

In the following, we report and discuss results by RQ.

The first and second rows of \tabref{tab:perfect} compare a non-pre-trained model (M0) with the same model pre-trained using a \emph{masked language model} self-supervised objective (M1, being a replica of VulRepair \cite{vulRepairFSE}, yet using our cleaned dataset without duplicates).


\rev{The achieved results clearly show the boost in performance that self-supervised pre-training provides for the task of vulnerability patching. To this extent, M0 patches vulnerabilities achieving a success rate ranging from 2.35\% to 4.79\%. These percentages correspond to the most challenging (\ie $K$=10) and favorable scenario for the model (\ie $K$=10). On the other hand, M1 demonstrates a superior ability to patch vulnerabilities, with a success rate of 3.34\% when recommending a single candidate patch (\ie $K=1$), which increases up to 6.90\% when the model suggests  $K$=10 candidate repairs.
}

Thanks to the self-supervised pre-training, M1 exhibits major improvements in terms of EMs for all values of $K$, with margins ranging from 3.34\% ($K$=1) to 6.90\% ($K=10$). Unsurprisingly, also the  CrystalBLEU exhibits an increase. According to Wilcoxon signed-rank test, the difference in terms of CrystalBLEU, is statistically significant ($p$-value), with a \emph{Large} Cliff’s Delta (d). We could not compute McNemar's test for EMs for most values of $K$ given the 0 EM predictions generated by the non-pre-trained model.

\begin{table}[b!]
	\centering
		\vspace{-0.4cm}
	\caption{Comparison in terms of percentage of Exact Match predictions reported by Fu \etal \cite{vulRepairFSE} and by our replication of their approach. The observed difference is due to our cleaning of their dataset which removed ``token-by-token'' overlapping instances between the training and the test set.}
		\label{tab:comparison}
			\resizebox{0.8\columnwidth}{!}{%
		\begin{tabular}{llrrrr}
			\toprule
			\multirow{2}{*}{\textbf{Source}}  & & \multicolumn{4}{c}{\textbf{Exact Matches (\%)}}\\\cline{3-6}
			& & {\bf K=1} & {\bf K=3} & {\bf K=5} & {\bf K=10}\\
			\midrule
			Fu \etal \cite{vulRepairFSE} & & 30\% & 38\% & 41\% & 42\%\\
			Our replication & & 3.34\% & 5.72\% & 6.54\% & 6.90\%\\\midrule
			& &  {\bf -26.66\%} & {\bf -32.28\%} & {\bf -34.46\%} & {\bf -35.10\%}\\
			\bottomrule
		\end{tabular}
	}
\scriptsize
\end{table}

\rev{When comparing the performances achieved on the cleaned VulRepair dataset, and those of the original paper by Fu \etal \cite{vulRepairFSE}, we observe a substantial drop in the ability of the model \cite{vulRepairFSE} to recommend candidates patches for vulnerable code. }
This is because the dataset of Fu \etal \cite{vulRepairFSE} has a significant overlap of instances between the training and the test set ($\sim$40\% of instances in the test set were present in the training set). \tabref{tab:comparison} reports the performance from the original paper (top row) as compared to our replication. As expected, the duplicated instances substantially inflated the EM predictions reported in \cite{vulRepairFSE}, with up to a +35.10\% for $K$=10. Nevertheless, as shown by our results on the cleaned dataset, we confirm that a large-scale self-supervised pre-training helps in boosting the performance of vulnerability patching.

\begin{resultbox}
	\textbf{Answer to RQ$_{1.1}$.} 
	Compared to techniques that do not rely on pre-training (\eg M0),  
	pre-trained models help to automatically fix a larger number of vulnerable functions, up to $\sim$2\% more. The non-pre-trained model struggles more, having a hard time learning the vulnerability patching task, due to the small fine-tuning dataset available.
\end{resultbox}

\subsection{RQ$_{1.2}$: How does the usage of supervised training affect the performances of large pre-trained models of code in automatically patching vulnerabilities?}

\vspace{0.1cm}

Seeding bug-fixing knowledge into the model (M2) as proposed by Chi \etal \cite{seqTrans} and Chen \etal \cite{vRepairTSE} provides a substantial boost in performance as compared to the use of self-supervised training alone (compare M2 to M1 in \tabref{tab:perfect}). The improvement is consistent across all beam sizes we experimented with. Specifically, when only relying on the top prediction (\ie $K$=1), the EM predictions increase from 3.34\% to 12.28\% (+$\sim$9\%). This gap becomes even higher when the model is asked to generate more candidate patches, with improvements in EM predictions of up to +$\sim$12\%.

\begin{figure}[t]
	\centering
	\includegraphics[width=0.9\columnwidth]{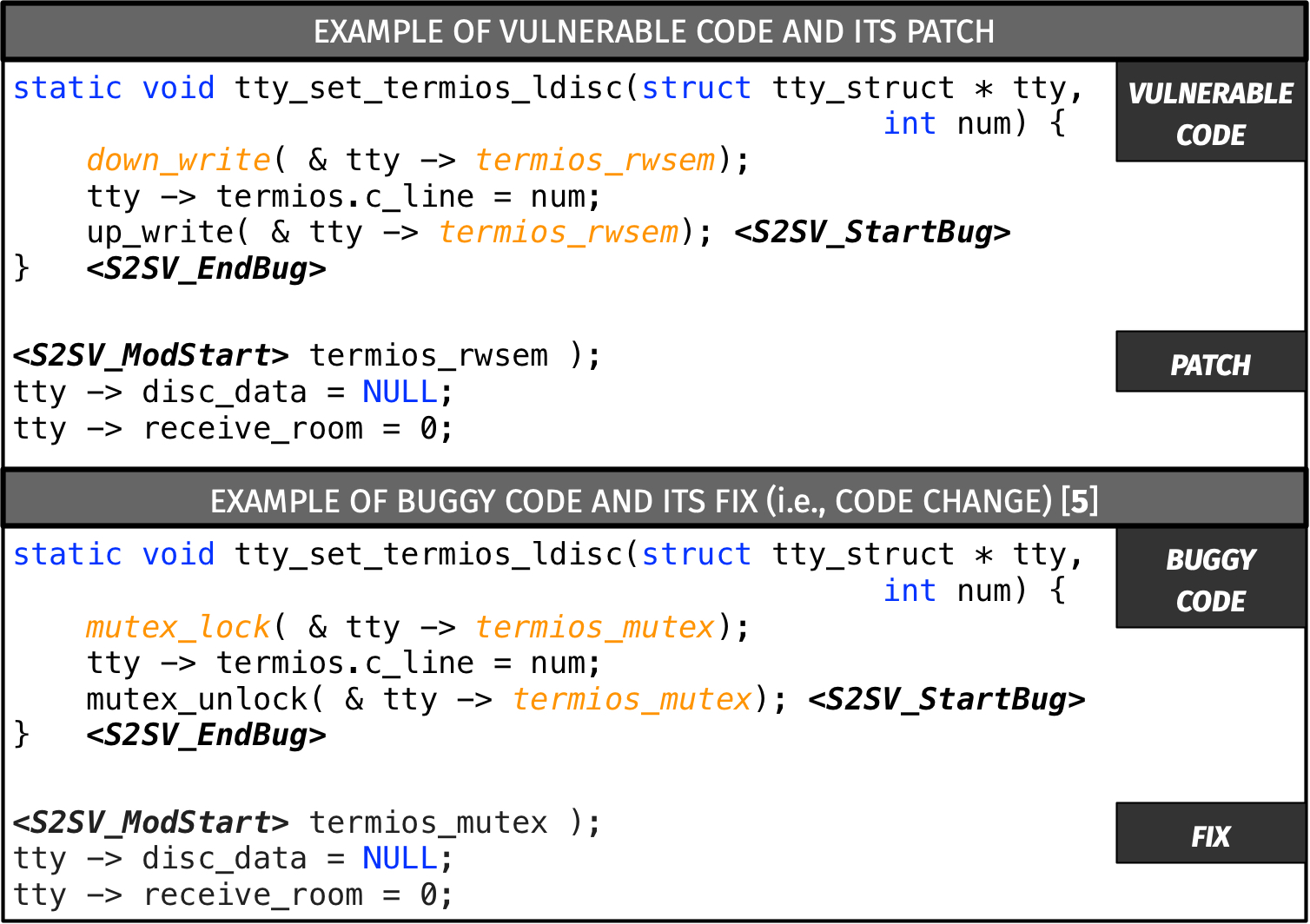}
	\caption{Example of vulnerable code vs. buggy code for which the model is required to perform similar code changes.}
	\vspace{-0.2cm}
	\label{fig:example}
	
\end{figure}

The superior performance ensured by the injection of bug-fixing knowledge is also confirmed by the McNemar test (see \tabref{tab:stats}), which reports a statistically significant difference in EM predictions between M1 and M2, with an OR=26.33 in favor of M2. The Wilcoxon signed-rank test also indicates a statistically significant difference ($p$-value $<$ 0.05) when comparing the distributions of CrystalBLEU scores between M1 and M2 with, however, a \emph{Negligible} effect size.
The usefulness of learning from bug fixes confirms what was found by previous literature \cite{vRepairTSE,seqTrans}. This is mainly due to the commonalities between a bug fix and a vulnerability fix. This helps the model to gain knowledge related to the downstream task without, however, using the scarce fine-tuning instances available for such a task. In the end, a vulnerability fix is nothing but a specific bug fix, where the bug could be possibly exploited for a security attack.
\figref{fig:example} shows a concrete example of such a scenario. The top part of the figure shows the changes needed to fix a vulnerability in our test set, while the bottom part shows a bug-fixing change from the dataset by Chen \etal \cite{vRepairTSE} (\ie the one used for the bug-fixing training). As it can be seen, while acting on different code components, both the involved functions require the same code changes. In other words, in both functions, the fields \texttt{disc\_data} and \texttt{receive\_room} need to be initialized to \texttt{NULL} and 0, respectively.
In such a case, the learned fixing pattern is basically the same. It is just that in one case the problem exposed the system to a possible vulnerability attack, in the other it did not, although it is also true that many bug fixes are silent vulnerability fixes \cite{zhou2021finding}.

\vspace{0.1cm}
\begin{resultbox}
	\textbf{Answer to RQ$_{1.2}$.} Incorporating task-specific knowledge via bug-fixing training into pre-trained models of code significantly increases the ability to generate patches for vulnerable code components. The increase in Exact Match predictions ranges between $\sim$9\% and $\sim$12\%, depending on the number of candidate solutions (\ie patches) generated by the models.
\end{resultbox}


\subsection{RQ$_{1.3}$: What is the role played by the prompt-tuning when producing patches for vulnerable code?}

We first focus on the effect of prompt-tuning when fine-tuning a model pre-trained using only a self-supervised procedure ($M3_{S_n}$ and $M3_{H_n}$ in \tabref{tab:perfect}). Afterward, we report and discuss the impact of prompt-tuning for a model pre-trained using both self-supervised and supervised (\ie bug-fixing) pre-training ($M4_{S_n}$ and $M4_{H_n}$ in \tabref{tab:perfect}).\smallskip

\vspace{-0.2cm}
\textbf{Prompt-tuning a self-supervised pre-trained model.} Both soft ($M3_{S_1}$ to $M3_{S_5}$) and hard ($M3_{H_1}$ to $M3_{H_5}$) prompt-tuning help in improving the performance of a model that was pre-trained using a self-supervised objective only. However, the magnitude of the improvement is substantially different. When the models are asked to generate a single prediction ($K$=1), soft prompt-tuning provides an increase in EMs between 2.14\% and 3.44\%. While these improvements may look minor, note that the baseline (\ie M1) achieves 3.34\% of EM predictions. Thus, the best-performing template in the context of prompt-tuning ($M3_{S5}$) more than doubles the predictive capabilities of the model. Remember that this template is the one providing the model with information about the type of vulnerability to fix: 
\texttt{This text <CWE\_NAME>  \{$cwe_{name}$\}  </CWE\_NAME> describes the vulnerable code \{$vulnerable_{code}$\}  fixed by:  \{$patch$\}}

The statistical tests (\tabref{tab:stats}) support such findings, with the models subject to soft prompt-tuning achieving, independently from the used prompt template, a statistically significant higher percentage of EM predictions than the self-supervised pre-trained models fine-tuned with a standard approach. The adjusted $p$-values are always $<$0.05, with ORs ranging from 2.51 (M1 \emph{vs} $M3_{S1}$) and 3.03 (M1 \emph{vs} $M3_{S5}$). 

Improvements are also observed in terms of CrystalBLEU score, which, in comparison with M1, has an increase of up to +12\% for $K=1$. Also in this case, the statistical tests confirm a significant difference in favor of the models fine-tuned with soft prompting, independently from the used template (in all cases, $p$-values $<$0.05 with a small effect size).
Looking at the different templates used for soft-prompting, we found them to have  a non-trivial impact on performance. When looking at the top-1 prediction ($K=1$), the best-performing template is, as said, $M3_{S5}$, while the worst one is $M3_{S1}$, with a 1.30\% gap in terms of EM predictions, which means a relative $\sim$24\% improvement (1.30/5.48). This points to the importance of carefully selecting the template to use when adopting prompt fine-tuning: The setting of the template must be considered as important as that of the other model's hyperparameters, that usually undergo a tuning procedure. It is also worth noticing that, while $M3_{S5}$ is the best in class for $K=1$, this is not the case for other $K$ values: for $K=3$ and $K=5$ it is the second-best, while for $K=10$ is the third-best, despite the minor gap from the best-performing template. Thus, even the usage scenario envisioned for the trained model and, consequently, the number of candidate solutions that it will be required to generate for each input may impact the choice of the template to adopt.  Looking at the hard prompting results ($M3_{H_1}$ to $M3_{H_5}$ in \tabref{tab:perfect}), the improvement it achieves compared to the baseline is  smaller than what we observed for soft-prompting (\eg +0.67\% when $K=1$). The lower (or, in a few cases, lack of) improvement holds for all experimented prompts, and for all the considered beam sizes $K$. Also, it is confirmed by the statistical tests, reporting non-significant $p$-values (see the comparison between M1 and $M3_{H_n}$ in  \tabref{tab:stats}).

\textbf{Prompt-tuning a self-supervised and supervised pre-trained model.} The results achieved when prompt-tuning a model that has already acquired  task-specific knowledge (thanks to the bug-fixing training) are quite different from what was achieved for the self-supervised pre-trained model. Looking at the $M4_{S_n}$ and $M4_{H_n}$ in \tabref{tab:perfect}, we can notice how, in this case, the models that have been subject to hard prompting are the ones working better and, overall, achieving the best performance in terms of EM predictions (see the values reported in bold faces). This holds for all $K$ values. The gap from the baseline (M2, which employs self-supervised + supervised training, without prompt-tuning) is relatively small ($\sim$+1\% in EMs) and statistically significant (see \tabref{tab:stats}). Note that this is in line with what was previously observed in the literature about the benefits of prompt-tuning when dealing with other software engineering  tasks \cite{wang2022no}. The soft prompt-tuning is, instead, able to improve the results in terms of CrystalBLEU, yet it has a price to pay in terms of EM predictions, which are worse than the M2 baseline, as also confirmed by the statistical analyses.

There may be two possible reasons why no further improvements are observed when using soft prompting on top of models that were subject to a combination of self-supervised and supervised training.

The first reason is related to the additional knowledge the model acquires from bug fixes. As previously shown (see \figref{fig:example}), fixing a generic bug may require code changes that are very similar (or even equal) to those needed when patching vulnerabilities. Therefore, the similarity between these tasks may already resemble code transformations for which the model has become proficient, and the effort in applying these transformations in a different context (\ie vulnerability fixing) is trivial. 

The second reason is related to prompt-tuning strategies in general, which heavily rely on natural language knowledge acquired during pre-training. This critical information may be partially overwritten when further specializing pre-trained models by incorporating context-specific knowledge. In other words, while the self-supervised pre-trained model primarily captures the statistical distribution of words in natural language, the model additionally trained for bug-fixing may have lost some of its ability to represent and interpret natural language, leading to poor performances.

That being said, prompt-tuning offers a notable advantage of potentially achieving high performance in fixing vulnerabilities without requiring the creation of a  bug-fixing dataset necessary for supervised training. 
Moreover, it is essential to consider the effort required for preprocessing these instances after collection, as well as the effort researchers must put into such a large-scale operation over a prolonged period.

\begin{resultbox}
	\textbf{Answer to RQ$_{1.3}$} Hard and (above all) soft prompt fine-tuning can be a relatively cheap way to boost performance for models only subject to self-supervised pre-training.
	The improvement introduced by prompting is less evident or negligible for models already having knowledge from similar tasks, \eg bug fixing.
\end{resultbox}

\vspace{-0.45cm}
\subsection{Implications of our findings}
\label{subsec:implication}

\revCR{Our findings have significant implications for both researchers and practitioners. Researchers should focus on overcoming the data scarcity challenges faced by generative models in vulnerability patching, while at the same time making sure of the quality of the collected elements used to train the proposed techniques, to avoid leakage of data, as shown in this research.
The promising initial results with prompt-tuning open up new opportunities for in-depth exploration of advanced methods for crafting effective prompts within the designated framework. Furthermore, we encourage researchers to develop new metrics that can effectively and accurately evaluate model performance in practical vulnerability patching scenarios.
On the other hand, practitioners need to recognize the limitations of these methods and avoid depending blindly on them. They should actively engage with researchers, providing targeted and practical feedback to refine and develop more sophisticated techniques with the goal in mind of further advancing patching vulnerabilities methods. This includes assessing generative solutions in real-world settings and identifying cases where they fall short. Such insights would guide the development of models, particularly when these solutions are employed to generate patches for critical environments such as, transportation networks, and healthcare systems. 
}

\section{Threats to Validity}
\label{sec:threats}

\textbf{Construct validity.} We used a consolidated set of measurements to assess the quality of vulnerability patching with respect to the dataset ground truth, \ie, percentages of perfect predictions, CrystalBLEU score \cite{eghbali2022crystalbleu}. Clearly, such a strategy might consider a wrong prediction legitimate fixes still differing from the ground truth.

\textbf{Internal validity.} Among the factors internal to our study that could influence our findings, the choice of DL model hyperparameters has a paramount role. To make the comparison fair, and since we leveraged a previous vulnerability patch dataset \cite{vulRepairFSE} and, therefore, we used the same hyperparameters used by VulRepair. Nevertheless, it is possible that prompt-tuning approaches or, in general, other configurations we experimented could work better with different unexplored settings. As explained in \secref{sec:ft-dataset}, we found duplicates in the VulRepair \cite{vulRepairFSE} dataset. To ensure a comparison on exactly the same test set used in the original paper, and considering that the training set is larger than the test set, we removed duplicates from the training set, leaving the test set unchanged. This, however, reduced the training material available for the model, compared to the original study. While a larger training set could produce better results, we have mitigated the threat by putting all considered treatments under the same conditions, \ie using the same training set. 

Our results (those of RQ$_{1.3}$ in particular) showed how soft prompt-tuning does not help for treatments (M4) where a supervised pre-training with bug fixes was performed, likely because the model lost the ``natural language'' knowledge in such a phase. This, however, also suggests that prompt-tuning could be done with bug fixes (and then followed by further fine-tuning with vulnerability patches). We tried a simple prompt-tuning with a subset of our data, but it did not produce significant improvements. Therefore, we abandoned such an option and did not include it in our results.

\textbf{Conclusion validity.} The study conclusions are supported by the use of suitable statistical procedures, namely a proportion test (McNemar \cite{mcnemar}) to compare perfect predictions, and Wilcoxon signed-ranked test \cite{wilcoxon} to compare distributions of CrystalBLEU scores. These tests are complemented by suitable effect size procedures, namely OR and Cliff's delta.

\textbf{External validity.} Although this study reported a wide set of possible combinations of training strategies (23 in total),  several, further dimensions could be investigated, including vulnerabilities in other programming languages, other language models (\eg large language models), and further prompt-tuning strategies, \eg automatically-generated ones \cite{nashidretrieval}. In this paper, we only experimented using the T5 pre-trained model, which is the one also used by VulRepair \cite{vulRepairFSE}. This allowed a fair comparison with the model they proposed (except for the removal of duplicates from the dataset). \revCR{That being said, with the recent surge of large language models such as GPT-3 \cite{brown2020language} and GPT-4 \cite{openai2023gpt4}, we cannot guarantee that the observed differences will remain consistent when the treatments are applied using models featuring an order of magnitude more parameters (\eg GPT-4).}

\vspace{-0.3cm}
\section{Conclusion}
\label{sec:conclusion}

This paper empirically compares different strategies to perform vulnerability fixing using generative models, comparing a total of 23 different treatments featuring self-supervised pre-training, supervised training (using a bug fix dataset as in a previous work~\cite{vRepairTSE}), and, last but not least, different templates for hard and soft prompt-tuning.

Results of the study indicate that: (i) unlike what was observed in a recent paper \cite{vulRepairFSE}, vulnerability fixing still benefits from supervised training on a similar task (\ie bug fixing) rather than just doing fine-tuning on a pre-trained model as observed by Chen \etal~\cite{vRepairTSE}; (ii) the use of prompt-tuning introduces benefits over self-supervised pre-trained models but also (limited to hard prompting and with a small magnitude) over models that also underwent supervised training and; (iii) while supervised training remains necessary to achieve the best performances, in the absence of a large dataset of bug fixes (which collection requires a non-trivial effort) or if no further computational resources can be invested in pre-training, the use of prompt fine-tuning constitutes a valid, cheap solution to boost a fine-tuning over a self-supervised pre-trained model. Future work aims at experimenting with large language models, as well as more complex or even automatically-generated prompting strategies such as those proposed by Nashid \etal \cite{nashidretrieval}.

\section{Data Availability}
\label{sec:replication}
Our replication package is available online \cite{replication}. Within it, we provide all code and data used in our study. This includes: (i) the datasets exploited for training/testing the experimented models; (ii) the code needed to train and test the models; and (iii) \texttt{R} scripts used for the statistical analysis. 

\section*{Acknowledgment}
This project has received funding from the European Research Council (ERC) under the European Union's Horizon 2020 research and innovation programme (grant agreement No. 851720). Massimiliano Di Penta acknowledges the Horizon 2020 (EU Commission) project COSMOS (DevOps for Complex Cyber-physical Systems), Project No. 957254-COSMOS.

\bibliographystyle{ACM-Reference-Format}
\bibliography{bibfile}

\end{document}